\documentstyle[aps,prd,tighten,epsf]{revtex}
\begin{document}
\draft
\newcommand{\be}{\begin{equation}}
\newcommand{\ee}{\end{equation}}
\newcommand{\ben}{\begin{eqnarray}}
\newcommand{\een}{\end{eqnarray}}

\newcommand{\la}{{\lambda}}
\newcommand{\Om}{{\Omega}}
\newcommand{\ta}{{\tilde a}}
\newcommand{\bg}{{\bar g}}
\newcommand{\bh}{{\bar h}}
\newcommand{\si}{{\sigma}}
\newcommand{\th}{{\theta}}
\newcommand{\C}{{\cal C}}
\newcommand{\D}{{\cal D}}
\newcommand{\cA}{{\cal A}}
\newcommand{\cT}{{\cal T}}
\newcommand{\cO}{{\cal O}}
\newcommand{\eeo}{\cO ({1 \over E})}
\newcommand{\G}{{\cal G}}
\newcommand{\cL}{{\cal L}}
\newcommand{\cH}{{\cal H}}
\newcommand{\cE}{{\cal E}}
\newcommand{\M}{{\cal M}}

\newcommand{\p}{\partial}
\newcommand{\na}{\nabla}
\newcommand{\ssum}{\sum\limits_{i = 1}^3}
\newcommand{\dssum}{\sum\limits_{i = 1}^2}
\newcommand{\tal}{{\tilde \alpha}}

\newcommand{\tp}{{\tilde \phi}}
\newcommand{\tPhi}{\tilde \Phi}
\newcommand{\tpsi}{\tilde \psi}
\newcommand{\tim}{{\tilde \mu}}
\newcommand{\tr}{{\tilde \rho}}
\newcommand{\tir}{{\tilde r}}
\newcommand{\rp}{r_{+}}
\newcommand{\hr}{{\hat r}}
\newcommand{\rv}{{r_{v}}}
\newcommand{\dr}{{d \over d \hr}}
\newcommand{\dR}{{d \over d R}}

\newcommand{\hhf}{{\hat \phi}}
\newcommand{\hhM}{{\hat M}}
\newcommand{\hhQ}{{\hat Q}}
\newcommand{\hht}{{\hat t}}
\newcommand{\hhr}{{\hat r}}
\newcommand{\hhS}{{\hat \Sigma}}
\newcommand{\hhD}{{\hat \Delta}}
\newcommand{\hhm}{{\hat \mu}}
\newcommand{\hro}{{\hat \rho}}
\newcommand{\hhz}{{\hat z}}

\newcommand{\tD}{{\tilde D}}
\newcommand{\tB}{{\tilde B}}
\newcommand{\tV}{{\tilde V}}
\newcommand{\hT}{\hat T}
\newcommand{\tF}{\tilde F}
\newcommand{\tT}{\tilde T}
\newcommand{\hC}{\hat C}
\newcommand{\ep}{\epsilon}
\newcommand{\bep}{\bar \epsilon}
\newcommand{\ppp}{\varphi}
\newcommand{\Ga}{\Gamma}
\newcommand{\ga}{\gamma}
\newcommand{\hth}{\hat \theta}
\title{Physical Process Version of the First Law of Thermodynamics
for Black Holes in Higher Dimensional Gravity}

\author{Marek Rogatko}
\address{Institute of Physics \protect \\
Maria Curie-Sklodowska University \protect \\
20-031 Lublin, pl.Marii Curie-Sklodowskiej 1, Poland \protect \\
rogat@tytan.umcs.lublin.pl \protect \\
rogat@kft.umcs.lublin.pl}
\date{\today}
\maketitle
\smallskip
\bigskip
\begin{abstract}
The problem of {\it physical process} version of the first law of black hole
thermodynamics for charged rotating black hole in $n$-dimensional gravity is elaborated.
The formulae for the first order variations of mass, angular momentum and canonical energy 
in Einstein $(n-2)$-gauge form field theory are derived. These variations are expressed by means of 
the perturbed matter energy momentum tensor and charge matter current density.
\end{abstract}
\baselineskip=18pt
\par
\section{Introduction}
In their seminal paper Bardeen, Carter and Hawking \cite{bar73} considering linear
perturbations of a stationary electrovac black hole to another stationary black hole
found the first law of black hole thermodynamics. 
Contrary to the derivation presented in Ref.\cite{bar73}
Sudarsky {\it et al.} \cite{sud92} using the Arnowitt-Deser-Misner (ADM)
formalism derived the first law of black hole thermodynamics
for Einstein-Yang-Mills (EYM) theory valid for arbitrary asymptotically flat perturbations of a stationary 
black hole. In Ref.\cite{rog98}
this method, was used to study the first law of black hole thermodynamics
in Einstein-Maxwell axion-dilaton gravity (EMAD) being the low energy limit
of the heterotic string theory. In Ref.\cite{rog05}
the first law of black hole mechanics in $n$-dimensional gravity was established. 
\par
The first law of black hole thermodynamics for an arbitrary diffeomorphism invariant Lagrangians 
with metric and matter fields possessing stationary and axisymmetric
black hole solutions was widely studied in Refs.\cite{wal}. The case of a charged and rotating black hole
where fields were not smooth through the event horizon was considered in Ref.\cite{gao03}.
The case of
the higher curvature terms and higher derivative terms in the metric was considered in
\cite{jac}, while a generalized theory of gravity subject to the Lagrangian being arbitrary function of metric,
Ricci tensor and a scalar field was treated in Ref.\cite{kog98}.\\
One can also think about a {\it physical process} version of the first law of black hole thermodynamics
obtained by changing a stationary black hole by some infinitisemal physical process, e.g.,
when matter is thrown into black hole. Assuming that the black hole eventually settle down to a 
stationary state and calculating the changes of black hole's parameters one can find this law.
If the resulting relation fails comparing to the known version of 
the first law of black hole thermodynamics it will provide inconsistency with the assumption
that the black hole settles down to a final stationary state. This fact will give
a strong evidence against cosmic censorship. The {\it physical process } version of the first 
law of black hole thermodynamics in Einstein theory was proved in \cite{wal94}. Then, it was
generalized for Einstein-Maxwell (EM) black holes in Ref.\cite{gao01} and for EMAD gravity black holes in
\cite{rog02}.\\
In our paper we shall find a {\it physical process} version of the first black hole mechanics
in Einstein $(n-2)$-gauge form field theory. The convention will follow Ref.\cite{wal84}.

\section{Physical process version of the first law of black hole mechanics}
We begin with the Lagrangian of generalized Maxwell $(n-2)$-gauge form field in
$n$-dimensional spacetime as follows:
\be
{\bf L } = {\bf \ep} \bigg(
{}^{(n)}R - F_{(n-2)}^2 \bigg),
\ee
where by $ {\bf \ep}$ we denote the volume element,
$F_{(n-2)} = dA_{(n-3)} $ is $(n-2)$-gauge form field.
One can remark that
using in $n$-dimensional spacetime the generalized Maxwell field
$F_{j_{1} \dots j_{n-2}}$ enables one to treat both {\it magnetic} and
{\it electric} components of it.
\par
Consider, next, 
the first order variation of the conserved quantities
in this theory. Our main task will be to obtain the explicit 
formulae for the variation of mass and angular momentum.
The result of the variations yields
\ben \label{dl}
\delta {\bf L} &=& {\bf \epsilon} \bigg(
G_{\mu \nu} - T_{\mu \nu}(F_{(n-2)}) \bigg)~ \delta g^{\mu \nu}
+ 2 (n - 2)! \bigg( \na_{j_{1}} F^{j_{1} \dots j_{n-2}} \bigg) 
\delta A_{j_{2} \dots j_{n-2}}
+ d {\bf \Theta},
\een
where the energy momentum tensor for $(n-2)$-gauge form field is given by the expression
\be
T_{\mu \nu}(F_{(n-2)}) =
(n - 2)  F_{\mu j_{2} \dots j_{n-2}} F_{\nu}{}^{j_{2} \dots j_{n-2}} - {1 \over 2}
g_{\mu \nu}  F_{j_{1} \dots j_{n-2}} F^{j_{1} \dots j_{n-2}}.
\ee
The totally divergent term in Eq.(\ref{dl}) is a functional of the field variables
$A_{j_{1} \dots j_{n-3}}$ and their variations $\delta A_{j_{1} \dots j_{n-3}}$
which for simplicity we have denoted respectively 
by $\psi_{\alpha}$ and $\delta \psi_{\alpha}$. Inspection of relation (\ref{dl}) 
reveals
the following
form of the symplectic $(n - 1)$-form
$\Theta_{j_{1} \dots j_{n-1}}[\psi_{\alpha}, \delta \psi_{\alpha}]$, namely
\be
\Theta_{j_{1} \dots j_{n-1}}[\psi_{\alpha}, \delta \psi_{\alpha}] =
\ep_{\mu j_{1} \dots j_{n-1}} \bigg[
\omega^{\mu} - 2 (n - 2)!~ F^{\mu j_{2} \dots j_{n-2}} \delta A_{j_{2} \dots j_{n-2}}
\bigg]
\ee
where $\omega_{\mu}$ implies
\be
\omega_{\mu} = \na^{\alpha} \delta g_{\alpha \mu} - \na_{\mu} 
\delta g_{\beta}{}{}^{\beta}.
\ee
By virtue of Eq.(\ref{dl}) one enables to read off
the source-free Einstein $(n-2)$-gauge form fields equations of motion
\ben \label{m1}
G_{\mu \nu} - T_{\mu \nu}(F_{(n-2)}) &=& 0, \\
\na_{j_{1}} F^{j_{1} \dots j_{n-2}} &=& 0.
\label{m2}
\een
One identifies the variations of the fields $\delta \psi_{\alpha}$
with a general coordinate transformations ${\cal L}_{\xi} \psi_{\alpha}$
induced by an arbitrary Killing vector $\xi_{\alpha}$. The
Noether $(n - 1)$-form with respect to this Killing vector $\xi_{\alpha}$ implies
\cite{wal}
\be
{\cal J}_{j_{1} \dots j_{n-1}} = \ep_{m j_{1} \dots j_{n-1}} {\cal J}^{m}
\big[\psi_{\alpha}, {\cal L}_{\xi} \psi_{\alpha}\big],
\ee
where the vector field ${\cal J}^{m}
\big[ \psi_{\alpha}, {\cal L}_{\xi} \psi_{\alpha}\big]$ is given by
\be
{\cal J}^{\delta}
\big[\psi_{\alpha}, {\cal L}_{\xi} \psi_{\alpha}\big] = \Theta^{\delta}
\big[\psi_{\alpha}, {\cal L}_{\xi} \psi_{\alpha}\big] - \xi^{\delta} L.
\label{curr}
\ee
Using the above relation (\ref{curr}) we get
the resultant expression for the Noether
current $(n - 1)$-form with respect to the Killing vector field $\xi_{\alpha}$.
It yields
\ben \label{ttt}
{\cal J}_{j_{1} \dots j_{n-1}} &=&
d Q_{j_{1} \dots j_{n-1}}^{GR} + 2 \ep_{\delta j_{1} \dots j_{n-1}}
\bigg(
G^{\delta}{}{}_{\eta} - T^{\delta}{}{}_{\eta}(F_{(n-2)})
\bigg) \xi^{\eta} \\ \nonumber
&-& 2(n - 2)! (n - 3) \ep_{m j_{1} \dots j_{n-1}} \na_{\alpha_{2}}
\bigg( \xi^{d}~ A_{d \alpha_{3} \dots \alpha_{n-2}} ~F^{m \alpha_{2} \dots \alpha_{n-2}} \bigg)\\ \nonumber
&+& 2(n - 2)! (n - 3)  \ep_{m j_{1} \dots j_{n-1}}
\xi^{k}~A_{k \alpha_{3} \dots \alpha_{n-2}}~ \na_{\alpha_{2}}
\bigg( F^{m \alpha_{2} \dots \alpha_{n-2}} \bigg),
\een
where $Q_{j_{1} \dots j_{n-2}}^{GR}$ implies
\be
Q_{j_{1} \dots j_{n-2}}^{GR} = - \ep_{j_{1} \dots j_{n-2} a b} \na^{a} \xi^{b}.
\ee
Quite non-trivial calculations reveal that the 
relation (\ref{ttt}) can be written in the following form:
\ben
{\cal J}_{j_{1} \dots j_{n-1}} &=&
d Q_{j_{1} \dots j_{n-1}} + 2 \ep_{\delta j_{1} \dots j_{n-1}}
\bigg(
G^{\delta}{}{}_{\eta} - T^{\delta}{}{}_{\eta}(F_{(n-2)})
\bigg) \xi^{\eta} \\ \nonumber
&+&
2 (n - 2)! (n - 3)  \ep_{m j_{1} \dots j_{n-1}} \na_{\alpha_{2}} F^{m \alpha_{2} \dots \alpha_{n-2}} 
~\xi^{k}~A_{k \alpha_{3} \dots \alpha_{n-2}}.
\een 
Having in mind that ${\cal J}[\xi] = d Q[\xi] + \xi^{\alpha}{\bf C_{\alpha}}$,
where  ${\bf C_{\alpha}}$ is an $(n - 1)$ form locally constructed from
the dynamical fields we may identify $Q_{\alpha \beta }$
as the Noether charge. Thus, the Noether charge is subject to the expression
\be
Q_{j_{1} \dots j_{n-2}} = Q_{j_{1} \dots j_{n-2}}^{GR} + Q_{j_{1} \dots j_{n-2}}^{F},
\ee
where
\be
Q_{j_{1} \dots j_{n-2}}^{F} = 2 (n - 3) \ep_{m k j_{1} \dots j_{n-2}}~F^{m k \alpha_{3} \dots \alpha_{n-2}} 
~\xi^{d}~A_{d \alpha_{3} \dots \alpha_{n-2}}.
\ee
On the other hand, the quantity $C_{a j_{1} \dots j_{n-1}}$ implies the following:
\be 
C_{a j_{1} \dots j_{n-1}} = 2 \ep_{m j_{1} \dots j_{n-1}}
\bigg[ G_{a}{}{}^{m} - T_{a}{}{}^{m}(F_{(n-2)}) \bigg] +
2 (n - 2)!~ (n - 3)~ \ep_{m j_{1} \dots j_{n-1}}~\na_{\alpha_{2}} F^{m \alpha_{2} \dots \alpha_{n-2}} 
A_{a \alpha_{3} \dots \alpha_{n-2}}.
\ee
If ${\bf C_{\alpha}}$ is equal to zero one has the source-free equations fulfilled, when it does not hold we obtain
\ben
G_{\mu \nu} - T_{\mu \nu}(F_{(n-2)}) &=& T_{\mu \nu}(matter) , \\
\na_{j_{1}} F^{j_{1} \dots j_{n-2}} &=& j^{j_{2} \dots j_{n-2}}(matter).
\een
Let us suppose that $(g_{\mu \nu},~ A_{\alpha_{1} \dots \alpha_{n-3}})$ be the solution of the 
source-free equations of motion with $(n-2)$-gauge form field. Let us assume further
that $(\delta g_{\mu \nu},~ \delta A_{\alpha_{1} \dots \alpha_{n-3}})$             
be linearized perturbations fulfilling equations of motion with
sources $\delta T_{\mu \nu}(matter)$ and $\delta j_{j_{2} \dots j_{n-2}}(matter)$.
Thus, for a perturbed $\delta C_{a j_{1} \dots j_{n-1}}$
quantity we have the following relation:
\be
\delta  C_{a j_{1} \dots j_{n-1}} = 2 \ep_{m j_{1} \dots j_{n-1}}
\bigg[ \delta T_{a}{}{}^{m}(matter) + (n - 2)!~ (n - 3)~ A_{a \alpha_{3} \dots \alpha_{n-2}}~ 
\delta j ^{m \alpha_{3} \dots \alpha_{n-2}}(matter) \bigg].
\ee
As was shown in Ref.\cite{gao01} when $\xi_{\alpha}$ is a Killing vector field of a background spacetime and it also
describes a symmetry of background matter fields. Then, one can write the explicit formula
for a conserved quantity $\delta H_{\xi}$ connected with the aforementioned Killing vector field, namely
\be
\delta H_{\xi} = - \int_{\Sigma} \xi^{\alpha} \delta {\bf C}_{\alpha}
+ \int_{\p \Sigma} \bigg(
\delta Q(\xi) - \xi \cdot \Theta \bigg).
\ee
In our case $\delta H_{\xi}$ has the form as follows:
\ben \label{hh}
\delta H_{\xi} &=& - 2 \int_{\Sigma}\ep_{m j_{1} \dots j_{n-1}} \bigg[
\delta T_{a}{}{}^{m}(matter) \xi^{a} + (n - 2)!~ (n - 3)~ \xi^{a} A_{a \alpha_{3} \dots \alpha_{n-2}}~ 
\delta j ^{m \alpha_{3} \dots \alpha_{n-2}}(matter) \bigg] \\ \nonumber
&+& \int_{\p \Sigma}\bigg[
\delta Q(\xi) - \xi \cdot \Theta \bigg].
\een
By virtue of
choosing $\xi^{\alpha}$ to be an asymptotic time translation $t^{\alpha}$
one can conclude that
$M = H_{t}$.
Thus, we finally obtain 
the variation of the ADM mass
\ben \label{mm}
\alpha~ \delta M &=& - 2 \int_{\Sigma} \ep_{m j_{1} \dots j_{n-1}} \bigg[
\delta T_{a}{}{}^{m}(matter) t^{a} + (n - 2)!~ (n - 3)~ t^{a} A_{a \alpha_{3} \dots \alpha_{n-2}}~ 
\delta j ^{m \alpha_{3} \dots \alpha_{n-2}}(matter) \bigg] \\ \nonumber
&+& \int_{\p \Sigma}\bigg[
\delta Q(t) - t \cdot \Theta \bigg],
\een
where $\alpha = {n-3 \over n-2}$.
On the other hand, taking the Killing vector fields $\phi_{(i)}$ which are responsible
for the rotation in the adequate directions, we arrive at the relations for angular 
momenta
\ben \label{jj}
\delta J_{(i)} &=& 2 \int_{\Sigma} \ep_{m j_{1} \dots j_{n-1}} \bigg[
\delta T_{a}{}{}^{m}(matter) \phi_{(i)}^{a} + (n - 2)!~ (n - 3)~ \phi_{(i)}^{a} A_{a \alpha_{3} \dots \alpha_{n-2}}~ 
\delta j ^{m \alpha_{3} \dots \alpha_{n-2}}(matter) \bigg] \\ \nonumber
&-& \int_{\p \Sigma}\bigg[
\delta Q({\phi}_{(i)} - {\phi}_{(i)} \cdot \Theta \bigg].
\een
In order to consider the {\it physical process} version of the first law of black hole thermodynamics
let us
suppose that one has a classical, stationary black hole solution to the equations of motion of Einstein
$(n-2)$-gauge form field system (\ref{m1})-(\ref{m2}). Then, one perturbs the considered black hole by
dropping charged matter. Assuming that the black hole will be not destroyed in the process
of this phenomenon and settles down to a stationary final state, one can calculate the change of black hole's
parameters. 
To proceed to the {\it physical process} version of the first law of black hole thermodynamics let us assume 
moreover
that
$(\delta g_{\mu \nu},~ \delta A_{\alpha_{1} \dots \alpha_{n-3}})$ are solutions to the source free
Einstein equations with $(n-2)$ form field. Furthermore, suppose that the event horizon
the Killing vector field $\chi^{\mu}$ is of the form as
\be
\chi^{\mu} = t^{\mu} + \sum_{i} \Omega_{(i)} \phi^{\mu (i)}
\ee
Let us assume further that $\Sigma_{0}$ is an asymptotically flat
hypersurface which terminates on the event horizon and take into account 
the initial data on $\Sigma_{0}$ for a linearized perturbations
$(\delta g_{\mu \nu},~ \delta A_{\alpha_{1} \dots \alpha_{n-3}})$
with $\delta T_{\mu \nu}(matter)$ and $\delta j^{\alpha_{2} \dots \alpha_{n-2}}(matter)$. We 
require that $\delta T_{\mu \nu}(matter)$ and $\delta j^{\alpha_{2} \dots \alpha_{n-2}}(matter)$
disappear
at infinity and the initial data for 
$(\delta g_{\mu \nu},~ \delta A_{\alpha_{1} \dots \alpha_{n-3}})$
vanish in the vicinity of the black hole horizon $\cal H$ on 
the hypersurface $\Sigma_{0}$. 
It envisages the fact that for the initial time
$\Sigma_{0}$, the considered black hole is unperturbed. 
Consequently,
taking into account Eqs.(\ref{mm}) and (\ref{jj}) and having in mind that the perturbations vanish near 
the internal boundary $\p \Sigma_{0}$ of the initial surface, we can write the following:
\ben \label{ppp}
\alpha~ \delta M &-&  \sum_{i} \Omega_{(i)} \delta J^{(i)} = \\ \nonumber
&-& 2 \int_{\Sigma_{0}} \ep_{m j_{1} \dots j_{n-1}} \bigg[
\delta T_{a}{}{}^{m}(matter) \phi_{(i)}^{a} + (n - 2)!~ (n - 3)~ \phi_{(i)}^{a} A_{a \alpha_{3} \dots \alpha_{n-2}}~ 
\delta j ^{m \alpha_{3} \dots \alpha_{n-2}}(matter) \bigg] \\ \nonumber
&=& \int_{\cH} \gamma ^{\alpha}~k_{\alpha}~\bep_{j_{1} \dots j_{n-1}},
\een
where $\bep_{j_{1} \dots j_{n-1}} = n^{\delta}~\ep_{\delta j_{1} \dots j_{n-1}}$ and
$n^{\delta}$ is the future directed unit normal to the hypersurface $\Sigma_{0}$.\\
In the last line of Eq.(\ref{ppp}) we replace $n^{\delta}$ by the vector $k^{\delta}$ tangent to the
affinely parametrized null generators of the black hole event horizon $\cH$. 
It can be done due
to the fact of the conservation of current $\gamma^{\alpha}$ and the assumption
that all of the matter falls into the considered black hole.
Of course, one should also
integrate over the event horizon $\cH$.
\par
Now, it is easy to see that we have left with the following:
\be
\alpha~ \delta M -  \sum_{i} \Omega_{(i)} \delta J^{(i)} 
+ \delta \cE_{F}
= 2 \int_{\cH}
\delta T_{\mu}{}{}^{\nu} \xi^{\mu} k_{\nu},
\ee
where by $\delta \cE_{F}$ is the canonical energy of $(n-2)$-gauge form fields \cite{rog05}.
\par
Our next task will be to find the change in the area of black hole horizon. Having in mind
 that the null generators of the event horizon of the perturbed black hole coincide with
the null generators of the unperturbed stationary black hole \cite{gao01} (when we use the diffeomorphism freedom in
identifying the perturbed spacetime with the background one)
the result of this gauge choice is that the perturbation in the location of the event horizon
disappears and we obtain that $\delta k_{\mu} \propto k_{\mu}$. 
\par
Consider, next the Raychauduri equation of the form as follows:
\be
{d \theta \over d \lambda} = - {\theta^{2} \over (n - 2)} - \sigma_{ij} \sigma^{ij}
- R_{\mu \nu} \xi^{\mu} \xi^{\nu}.
\ee
Because of the fact that
the expansion $\theta$
and shear $\sigma_{\mu \nu}$ vanish in the stationary background, one gets
the perturbed Raychauduri's equation written as
\be
{d(\delta \theta) \over d \lambda} =
- \delta \bigg(
T_{\mu \nu}(total) k^{\mu} k^{\nu} \bigg) \mid_{\cH} =
- \delta \bigg(
T_{\mu \nu}(matter) \bigg)
 k^{\mu} k^{\nu}\mid_{\cH} 
- \delta \bigg(
T_{\mu \nu}(F_{(n-2)}) \bigg) 
k^{\mu} k^{\nu} \mid_{\cH},
\label{ray}
\ee 
where we exploit the fact that 
$T(F_{(n-2)})_{\mu \nu} k^{\mu}k^{\nu} \mid_{\cH} = 0$ and $\delta k_{\mu}
\propto k_{\mu}$ to eliminate terms in the form
$T(F_{(n-2)})_{\mu \nu} k^{\mu} \delta k^{\nu}$.
One finds that the remaining term in Eq.(\ref{ray}) reads 
\be
\delta T_{\mu \nu}(F_{(n-2)})~ k^{\mu} k^{\nu} \mid_{\cH} =
\bigg[ 2(n - 2) \delta F_{\mu j_{2} \dots j_{n-2}} F_{\nu}{}{}^{j_{2} \dots j_{n-2}} 
- {1 \over 2} \delta g_{\mu \nu} F_{j_{1} \dots j_{n-2}}F^{j_{1} \dots j_{n-2}}
- g_{\mu \nu} \delta  F_{j_{1} \dots j_{n-2}}~ F^{j_{1} \dots j_{n-2}}
\bigg]  k^{\mu} k^{\nu}.
\ee 
The last two expressions in the above equation are equal to zero, because 
of the fact that the vector $k_{\mu}$
is a null vector both in the perturbed as well as in unperturbed case. One also finds that
$F_{\nu j_{2} \dots j_{n-2}}  k^{\nu} \propto k_{j_{2}} \dots k_{j_{n-2}}$. 
For since we
take into account the antisymmetricity of 
$\delta F_{\mu j_{2} \dots j_{n-1}}$ the first term in the related equation also vanishes.\\
Hence, one concludes that
\be
{d(\delta \theta) \over d \lambda} =
- \delta
T_{\mu \nu}(matter)
 k^{\mu} k^{\nu}\mid_{\cal H},
\label{aaa}
\ee
Calculations of the right-hand side of Eq.(\ref{aaa}) are identical as in Ref.\cite{wal94}. For the readers'
convenience we quote the main steps. Namely, it is possible to
substitute for the Killing vector 
 $k_{\mu}$
the following expression 
\be
k_{\mu} = \bigg(
{\p \over \p V} \bigg)_{\mu} = {1 \over \kappa V} \bigg(
t^{\mu} + \sum_{i } \Omega_{(i)} \phi^{\mu (i)}
\bigg),
\ee
where $\kappa$ is the surface gravity. On the other hand, $V$ is
an affine parameter
along the null geodesics tangent to $\xi_{\beta}$
generating the adequate Killing horizon.
One can introduce the function $v$ ( it is called {\it Killing parameter time})
on the portion of Killing horizon. It
satisfies the relation
$\xi^{\beta} \na_{\beta} v = 1$ and it is related with $V$ by
the expression $V = exp(\kappa v)$.
Then, we multiply both sides of the resulting equation by $\kappa V$
and integrate over the event horizon.
\par
One should also
recall that expansion $\theta$ measures the local rate of change of 
the cross-sectional area as the observer moves up the null geodesics.
It can be parameterize  in the following way:
$\theta = {1 \over A}{d A \over d \lambda}$, where $\lambda$ is an
affine parameter which parametrized null geodesics generators of the 
horizon. 
The left-hand side of Eq.(\ref{aaa}) is evaluated by integration by parts, 
having in mind that $V = 0$ at the lower limit and $\theta$ has to vanish
faster than $1/V$ as $V$ tends to infinity when the considered black hole
settled down to a stationary final state after throwing some charged matter
into it.\\  
The consequence of the above establishes the result
\be
\kappa~ \delta A = \int_{\cal H} \delta
T^{\mu}{}{}_{\nu}(matter) \xi^{\nu} k_{\mu}.
\ee
In the light of what has been shown we obtained the
{\it physical process} version of the first law of black hole
mechanics in Einstein $(n-2)$-gauge form fields gravity of the same form as known from Ref.\cite{rog05}, namely
\be
\alpha~ \delta M - \sum_{i} \Omega_{(i)} \delta J^{(i)} 
+ \delta \cE_{F}
 = \kappa ~\delta {\cal A}.
\ee
Like in four-dimensional case a proof of the {\it physical process} version of the first law of
thermodynamics for $n$-dimensional black hole also provides support for cosmic censorship.



\begin{references}
%
\def\cmp#1#2#3{{ Commun. Math. Phys.} {\bf #1}, #2 (#3)}
\def\lmp#1#2#3{{ Lett. Math. Phys.} {\bf #1}, #2 (#3)}
\def\hpa#1#2#3{{ Hell. Phys. Acta} {\bf #1}, #2 (#3)}
\def\grg#1#2#3{{ Gen. Rel. Grav.} {\bf #1}, #2 (#3)}
\def\pr#1#2#3{{ Phys. Rev.} {\bf #1}, #2 (#3)}
\def\prl#1#2#3{{ Phys. Rev. Lett.} {\bf #1}, #2 (#3)}
\def\prd#1#2#3{{ Phys. Rev. D} {\bf #1}, #2 (#3)}
\def\pl#1#2#3{{ Phys. Lett} {\bf #1}, #2 (#3)}
\def\pla#1#2#3{{ Phys. Lett. A} {\bf #1}, #2 (#3)}
\def\plb#1#2#3{{ Phys. Lett. B} {\bf #1}, #2 (#3)}
\def\prep#1#2#3{{ Phys. Reports} {\bf #1}, #2 (#3)}
\def\phys#1#2#3{{ Physica} {\bf #1}, #2 (#3)}
\def\jcp#1#2#3{{ J. Comput. Phys.} {\bf #1}, #2 (#3)}
\def\jmp#1#2#3{{ J. Math. Phys.} {\bf #1}, #2 (#3)}
\def\jpm#1#2#3{{ J. Phys. A: Math. Gen.} {\bf #1}, #2 (#3)}
\def\cpr#1#2#3{{ Computer Phys. Rept.} {\bf #1}, #2 (#3)}
\def\cqg#1#2#3{{ Class. Quantum Grav.} {\bf #1}, #2 (#3)}
\def\cma#1#2#3{{ Computers Math. Applic.} {\bf #1}, #2 (#3)}
\def\mc#1#2#3{{ Math. Compt.} {\bf #1}, #2 (#3)}
\def\apj#1#2#3{{ Astrophys. J.} {\bf #1}, #2 (#3)}
\def\apjs#1#2#3{{ Astrophys. J. Suppl.} {\bf #1}, #2 (#3)}
\def\acta#1#2#3{{ Acta Astronomica} {\bf #1}, #2 (#3)}
\def\apl#1#2#3{{Ann. Physik. (Leipzig)} {\bf #1}, #2 (#3)}
\def\sa#1#2#3{{ Sov. Astro.} {\bf #1}, #2 (#3)}
\def\sia#1#2#3{{ SIAM J. Sci. Statist. Comput.} {\bf #1}, #2 (#3)}
\def\aa#1#2#3{{ Astron. Astrophys.} {\bf #1}, #2 (#3)}
\def\mnras#1#2#3{{ Mon. Not. R. astr. Soc.} {\bf #1}, #2 (#3)}
\def\npb#1#2#3{{ Nucl. Phys. B} {\bf #1}, #2 (#3)}
\def\prsla#1#2#3{{ Proc. R. Soc. London, Ser. A} {\bf #1}, #2 (#3)}
\def\jhep#1#2#3{{ JHEP} {\bf #1}, #2 (#3)}
\def\nuc#1#2#3{{Nuovo Cimento B } {\bf #1}, #2 (#3)}
\def\mpl#1#2#3{{ Mod. Phys. Lett. A} {\bf #1}, #2 (#3)}

\def\hepth#1#2{{ hep-th }{\bf #1} (#2)}
\def\grqc#1#2{{ gr-qc }{\bf #1} (#2)}
%
\bibitem{bar73}
J.M.Bardeen, B.Carter and S.W.Hawking, \cmp{31}{161}{1973}.
\bibitem{sud92}
D.Sudarsky and R.M.Wald, \prd{46}{1453}{1992}.
\bibitem{rog98}
M.Rogatko, \prd{59}{104010}{1998}.
\bibitem{rog05}
M.Rogatko, \prd{71}{024031}{2005}.
\bibitem{wal}
R.M.Wald, \prd{48}{R3427}{1993},\\
V.Iyer and R.M.Wald, \prd{50}{846}{1994},\\
V.Iyer and R.M.Wald, \prd{52}{4430}{1995},\\
V.Iyer, \prd{55}{3411}{1997}.
\bibitem{gao03}
S.Gao, \prd{68}{044016}{2003}.
\bibitem{jac}
T.Jacobsen, G.Kang and R.C.Myers, \prd{49}{6587}{1994},\\
T.Jacobsen, G.Kang and R.C.Myers, \prd{52}{3518}{1995}.
\bibitem{kog98}
J.Koga and K.Maeda, \prd{58}{064020}{1998}.
\bibitem{wal94}
R.M.Wald, {\it Quantum Field Theory in Curved Spacetime and Black Hole Thermodynamics}, 
University of Chicago Press (Chicago, 1994).
\bibitem{gao01}
S.Gao and R.M.Wald, \prd{64}{084020}{2001}.
\bibitem{rog02}
M.Rogatko, \cqg{19}{3821}{2002}.
\bibitem{wal84}
R.M.Wald, {\it General Relativity}, University of Chicago Press (Chicago, 1984).





\end{references}
\end{document}